# Deterministic Quantum Dot Cavity Placement Using Hyperspectral Imaging with High Spatial Accuracy and Precision.

Quirin Buchinger[1, *], Constantin Krause[1], Aileen Zhang[1,2], Giora Peniakov[1], Mohamed Helal[1], Yorick Reum[1], Andreas Theo Pfenning[1], Sven Höfling[1], and Tobias Huber-Loyola[1]

[1] Julius-Maximilians-Universität Würzburg, Physikalisches Institut, Lehrstuhl für Technische Physik, Am Hubland, 97074 Würzburg, Deutschland
[2] University of Arizona, Wyant College of Optical Sciences, 1630 E University Blvd, Tucson Arizona, USA
 * Corresponding author: Quirin Buchinger

Tel.: +49 931 31-87907
E-mail address: quirin.buchinger@uni-wuerzburg.de

**Abstract**

Single emitters in solid state are great sources of single and entangled photons. To boost their extraction efficiency and tailor their emission properties, they are often incorporated in photonic nanostructures. However, achieving accurate and reproducible placement inside the cavity is challenging but necessary to ensure the highest mode overlap and optimal device performance. For many cavity types —such as photonic crystal cavities or circular Bragg grating cavities — even small displacements lead to a significantly reduced emitter-cavity coupling. For circular Bragg grating cavities, this yields a significant reduction in Purcell effect, a slight reduction in efficiency and it introduces polarization on the emitted photons. Here we show a method to achieve high accuracy and precision for deterministically placed cavities on the example of circular Bragg gratings on randomly distributed semiconductor quantum dots. We introduce periodic alignment markers for improved marker detection accuracy and investigate overall imaging accuracy achieving (9.1 ± 2.5) nm through image correction. Since circular Bragg grating cavities exhibit a strong polarization response when the emitter is displaced, they are ideal devices to probe the cavity placement accuracy far below the diffraction limit. From the measured device polarizations, we derive a total spatial process accuracy of (33.5 ± 9.9) nm based on the raw data, and an accuracy of (15 ± 11) nm after correcting for the system response, resulting in a device yield of 68 % for well-placed cavities.

Key words: Quantum Dot, Circular Bragg Grating, QD Imaging, Hyperspectral Imaging, Deterministic Fabrication, Single-Photon Source, Photoluminescence Imaging

# 1. Introduction

The generation of on-demand single photons or entangled photons is a key requirement for quantum information technologies, including quantum communication, quantum computing, and quantum metrology [1–5]. Semiconductor quantum dots (QDs) have emerged as one of the most promising platforms as non-classical light sources for these applications, offering high brightness, high-fidelity entanglement, and near-zero multi-photon events [6, 7]. One key advantage of QDs over other single-photon sources is their capability for deterministic photon generation. Unlike probabilistic sources, where single photon emission is stochastic, a resonantly excited QD emits a single photon per excitation cycle, achieving near-unity internal quantum efficiency [8]. Furthermore, the emission wavelength can be tailored through material composition, growth conditions and strain engineering [9], enabling integration into quantum photonic circuits [10, 11]. Beyond single-photon emission, quantum dots also serve as highly efficient entangled photon-pair sources. When a quantum dot undergoes the biexciton-exciton cascade, the emitted photon pair can exhibit polarization entanglement [12]. Besides the biexciton-exciton cascade, the spin of an electron or hole in a QD trion state, controlled via a suitable excitation sequence and an external magnetic field, can serve as an entangler to generate a chain of entangled photons [13].

Among the challenges of using QDs is that as grown, in a bulk-like structure, most of the emitted light is constrained in the semiconductor by total internal reflection. To increase the extraction efficiency and further enhance the QD brightness by using the Purcell effect, the QDs are embedded in photonic nanostructures such as micropillar cavities [8], photonic crystals cavities [14] or circular Bragg grating cavities [15–19]. The latter two of these cavity types usually have very small mode volumes with lateral mode sizes in the order of 100 nm, which makes the placement accuracy of the QD inside the cavity more demanding as compared to micropillar cavities, which have a lateral mode expansion on the micrometer scale. In addition, typically used QDs, grown by self-assembled or droplet etching techniques, exhibit random positioning and variations in size and composition, leading to an emission wavelength distribution [20]. This necessitates precise QD localization and selection techniques. To tackle this challenge different techniques for deterministic integration of QDs in photonic nanostructures have been developed [21]. Among the earliest approaches was in-situ photolithography [22]. This was later followed by marker-based photoluminescence (PL) imaging [17, 23], cathodoluminescence imaging [24], and in-situ electron beam lithography (EBL) [25]. More recently, a snapshot hyperspectral imaging approach for emitters embedded in one-dimensional planar cavities has been introduced [26]. Different studies using these approaches address the localization accuracy of QDs or alignment marker localization accuracy [17, 23, 24, 27] achieving QD localization accuracies down to 5 nm [28]. However, as discussed in ref. [24] additional sources of uncertainty, e.g. image distortion [27] must be considered to assess the total alignment accuracy. Ref. [24] compares different marker based PL imaging configurations and compares them with marker based and in-situ EBL. They report total positioning accuracy down to $(24 \pm 34)$ nm for cathodoluminescence imaging by measuring the QD position inside circular mesa structures using cathodoluminescence imaging.

To our knowledge, there is no study that statistically evaluates the positioning accuracy based on actual device performance. Here, we use a combination of hyperspectral and photoluminescence imaging to fabricate 87 circular Bragg grating cavities (CBGs). We introduce a periodic alignment marker designed to improve marker localization accuracy. Additionally, we investigate and discuss QD localization accuracy by analyzing alignment marker test fields, considering not only marker and QD localization accuracy but also additional uncertainties such as image distortion. Finally, we carry out polarization-resolved µPL measurements on the fabricated devices and infer the QD displacement from the measured device polarization [29]. The investigated circular Bragg grating cavity is the ideal vehicle to probe displacement since the device polarization allows for a determination of the positioning much better than the resolution limit. In contrast to photonic crystal cavities the device has cylindric symmetry, which results in a rotation of the polarization for different directions of displacement and the magnitude of the polarization measures the displacement. In principle super resolution techniques would be superior in resolution [30–32]. However, they would also suffer from image distortion, which is our main source of errors. Furthermore, fitting QD locations is already very precise and realizing markers out of a material that can be super-resolved seems to be challenging, since the marker also needs to be detectable with an electron beam for the EBL overlay.

## 2. Methods

We use a photoluminescence imaging technique to determine the position of QDs with respect to alignment markers similar to the technique reported in Ref. [17] extended with a scanning hyperspectral imaging routine to assess the emission wavelengths in every pixel of the image. The sample consists of a gold reflector followed by a dielectric spacer layer (out of $SiO_2$) and a GaAs membrane with Stranski-Krastanov grown InAs QDs embedded at a height, where the field maximum was calculated. On top of the surface, we place alignment markers for orientation on the sample which were fabricated using EBL and subsequent gold deposition. The alignment markers span fields of 37 x 37 µm$^2$. The sample is mounted in a closed cycle cryostat at a base temperature of $T = 4$ K. The sample is excited with a 730 nm continuous-wave (CW) laser diode. By focusing the laser on the back-focal plane of a 50x cover glass corrected objective with a N.A. of 0.65 the total field of view is excited at once. Additionally, for reflection imaging the sample surface is illuminated with an LED around 940 nm (see Fig. 1). To achieve optimal spatial accuracy while capturing the full wavelength information of all the QDs, the measurement routine is divided into two parts: a hyperspectral imaging scan and a single snapshot photoluminescence image.

### 2.1 Hyperspectral Imaging scan
To obtain the wavelength information of a full field, i.e., the sample space between the alignment markers, a real space image of the sample is projected onto the entrance port of a 500 mm Czerny-Turner spectrometer using a 4-f setup, see Fig. 1. By narrowing the slit to 50 µm, a single strip of the real space image is selected. The light emitted from the real space locations within this strip is analyzed using the spectrometer grating. As a result, each pixel row of the camera sensor corresponds to a position along the selected strip and all columns in this row contain the corresponding spectra. By moving the lens stepwise in front of the spectrometer, the real space image of the sample surface is shifted over the entrance slit. Thus, a different strip of locations of the sample is measured. These images, containing real space vs. wavelength information, are combined into a 3D data cube that contains both real space dimensions (in pixels) and the wavelength information of each pixel as a third dimension. Based on this data cube, the most suitable QDs for device fabrication are selected. To be selected the QDs should have narrow, high luminescent lines in the target wavelength range and no nearby visible neighbors. Due to the scanning approach and the additional optical elements inside the spectrometer, like the curved mirrors, which are optimized for spectral resolution and not for imaging accuracy and the off-axis grating, the hyperspectral image shows significant aberrations. These aberrations are so severe in our case that we could not correct them using imaging correction algorithms. Thus, we introduce a second step: photoluminescence imaging.

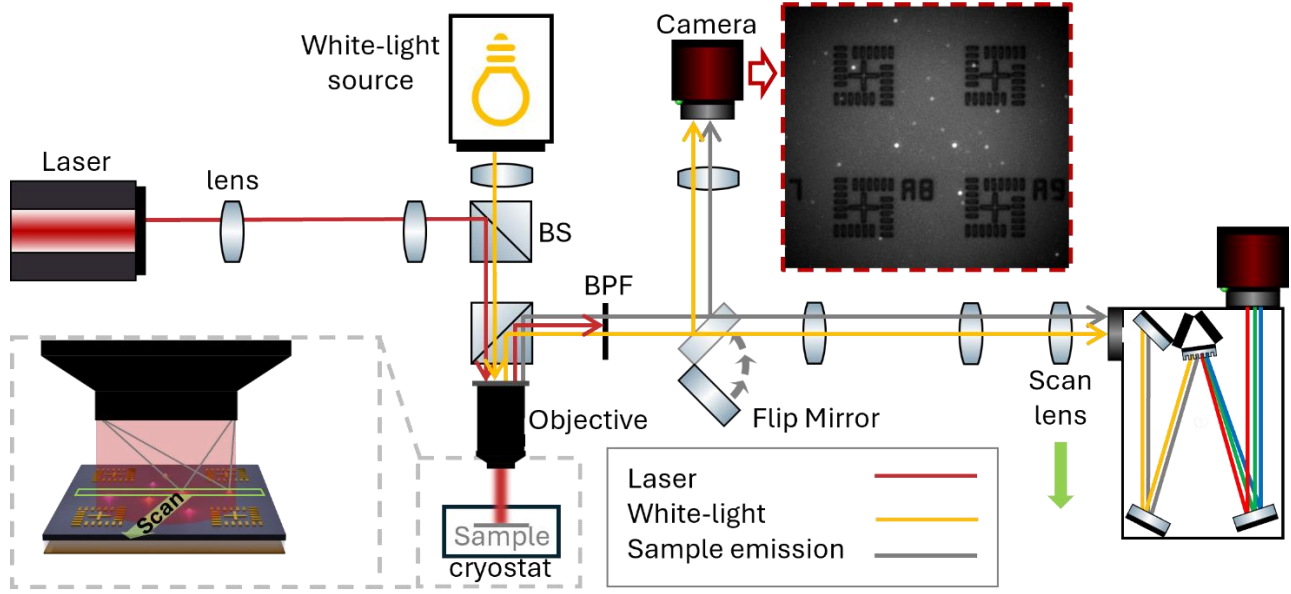

**Figure 1.:** Schematic of the hyperspectral imaging setup. A 730 nm continuous-wave (CW) laser is focused onto the back focal plane of the objective to simultaneously excite a large area of the sample. A 940 nm LED provides illumination for the sample surface and markers. The emitted and reflected light is filtered through a band-pass filter (BPF) before being projected onto the entrance slit of the spectrometer. A scan lens, mounted on a motorized linear stage, is used to focus different parts of the photoluminescence emission onto the entrance slit, enabling to analyze strip by strip of the sample image. A second camera, integrated via a flip mirror, acquires photoluminescence (inset) and reflection images to localize the quantum dots and markers.

## 2.2 Photoluminescence imaging

To reduce the number of optical elements and thereby minimize image distortions, a second camera is introduced into the setup, see Fig. 1. Using this camera, we can acquire fast images with minimal distortions to assess the position of the QDs. The wavelength range of interest, 930 nm, is selected using a 10 nm FWHM band-pass filter. Two images of the same field are acquired back-to-back: a photoluminescence image with switched on laser and minimal LED illumination, see red-dashed inset in Fig. 1, and one reflection image with high LED power and switched off laser, see Fig. 2 a). The first image captures the QD emission, while the second image clearly shows the markers, ensuring accurate marker fitting. We capture two separate images as the emission of a QD beneath an alignment marker changes the intensity profile along a marker and thus reduces the ability to accurately localize it in our fit routine. To reduce image distortion, an image correction method based on Zhang's method [24] is applied to both images. The markers in the reflection image are fitted (see section 2.3), and their positions are translated to the photoluminescence image to establish a coordinate system. Here the QD positions are fitted using a 2D Gaussian function.

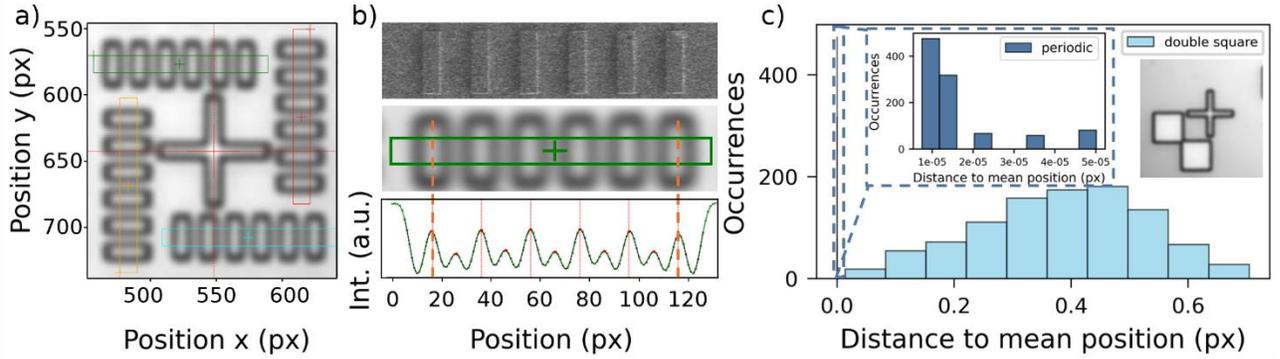

**Figure 2.: (a)** Zoomed-in reflection image of a periodic marker composed of four periodic structures and a central cross. The cross is used in the electron beam lithography (EBL) system to localize the marker position. To determine the marker position in the reflection image, the intensity along the four periodic features is binned and fitted. Panel **(b)** compares a scanning electron microscope (SEM) image, a reflection image, and the binned intensity profile of one quarter of the marker. The binned region is highlighted (green). The intensity is fitted using a multi-Gaussian fit with fixed spacings. **(c)** Test of the marker fitting routine by fitting the same double-square (inset) and periodic marker 1000 times with random variations in the starting position. The mean position of all fits is calculated, and the distance from each fit to the mean position is plotted. For the periodic marker, the distances are orders of magnitude smaller.

**2.3 Alignment marker localization**

The concept behind determining the position of the alignment markers is to fit the intensity profiles over features in the reflection image. The basic variant is to use two features per x- and y-direction, like the four legs of a cross [17, 24, 28] or the steps over the edges of two squares meeting at one corner (double square marker, compare Fig. 2 c) inset). We introduce markers with two periodic structures per x- and y-direction, each consisting of six gold strips surrounding a cross (periodic marker, see Fig. 2 a)). The cross is needed on our EBL machine for the overlay of the cavity structure. We fit a section of each periodic structure with a multiple Gaussian fit with fixed spacing (Fig. 2 b)) to get the position of the marker. We expected the periodic structures to increase the localization accuracy by increasing the number of features $n$ by a factor of six compared to a cross and thus reducing the error $\propto 1/\sqrt{n}$. However, due to fit constraints on the relative distance the fit error is reduced much more than this, see Fig. 2 c) and the results section for a detailed explanation.

.

## 2.4 Circular-Bragg Grating fabrication.

The determined QD positions are translated to coordinates for the EBL system. For each individual selected QD, a circular Bragg grating tailored to its emission wavelength by adjusting the center disk diameter is defined. The CBGs are patterned on the sample using the EBL. During this process, the EBL scans the crosses of the alignment markers to establish a reference coordinate system. The CBGs are structured by chemical dry etching using an inductively coupled Cl/Ar plasma.

## 2.5 Alignment analysis by polarization resolved measurements

If an emitter within a CBG cavity is not centered, its emission will be polarized [29]. This allows you to draw conclusions about the positioning of a QD in the device. To measure the device polarization a half-wave and quarter-wave plate together with a linear polarizer placed above the objective are used. By adjusting the waveplates accordingly, the QD emission under 730 nm CW laser excitation is recorded for six polarization states (H, V, D, A, R, L) using a spectrometer. As the laser passes through the waveplates as well, the excitation must not transfer its polarization to the QDs, as this would lead to an unexpected bias in the photoluminescence analysis. The used wavelength, which is above the GaAs band gap, ensures a polarization scrambling, which can be seen by the reference measurements, see Fig. 3 grey datapoints. The exact point from which the signal of a CBG is collected can influence the measured polarization [33]. To overcome this, the spectrometer slit is increased to 750 µm. Thus, the emission of the whole device is collected but the spectral resolution is reduced. The intensity for all emission wavelengths where the intensity of at least one polarization projection is above an arbitrary (but constant) threshold is summed up and the Stokes parameter S1, S2 and S3 are evaluated. Since the circular degree of polarization S3 shows zero, within the measurement uncertainty, as expected from the simulations, we consider the linear degree of polarization $\Pi = \sqrt{S_1^2 + S_2^2}$. Using the simulation from ref [29] this degree of polarization is translated to displacement. As a reference, we collected data from QDs outside the processed cavities, which should have no degree of polarization. The width of the reference signal cannot be explained by shot noise and may be explained by imperfect wave plates. The reference signal is fitted with a Nakagami distribution and taken as the system response (SRF). To correct the measured displacement with the SRF, the data is fitted with a convolution of SRF and a Nakagami distribution. Since zero displacement cannot be resolved even in the reference measurement, we account for this offset by shifting the x-axis range of the SRF such that its peak aligns with its mean value, rather than the center of the axis, before performing the deconvolution. This approach ensures that the corrected distribution reflects the true displacement relative to the system's intrinsic offset.

## 3. Results and Discussion

### 3.1. Periodic marker localization precision and accuracy

To get a better understanding of the reliability of the marker localization routine we repeatedly fitted the position of one double square marker (one periodic alignment marker) 1000 times in the same image using different initial guesses for the fit routines starting positions. We then calculated the distance of each result to the mean fit position of all fits. The result is shown in Fig 2. c). For the double square markers, we get a mean distance of $(0.38 \pm 0.14)$ px or $(38 \pm 14)$ nm, as one pixel corresponds to $(99.56 \pm 0.08)$ nm. For the periodic marker the distance to the mean position is $(2 \pm 1) *10^{-5}$ px, or $(2 \pm 1) *10^{-3}$ nm. This shows that our fitting routine is robust, reproducible and satisfying the required precision. However, the marker fitting precision does not reflect the marker fitting accuracy at all, as just one marker from one image is considered multiple times. To quantify the periodic marker fitting accuracy, marker test fields are investigated. These marker test fields (see Fig 4. inset) have nine markers with a distance of 18.5 µ$m$. One marker test field is measured 500 times consecutively and all the markers are fitted. Due to vibrations of the cryostat cold head, the position of the same marker shifts back and forth by $(6.07 \pm 0.67)$ nm in consecutive measurements. However, the distance between two neighbored markers should stay the same. We do observe a change of distance between two adjacent markers between two consecutive measurements as (2.87 ± 0.65) nm, which is our true marker fitting accuracy. Compared to previously reported marker location uncertainties, which sometimes reflect precision, the accuracy in marker fitting is increased by a factor of about 1.5 to 6 [17, 24]. A robust marker fitting routine is essential to achieve sufficient spatial accuracy as the markers are used to set up a coordinate system in which the QD are localized. As shown here on these periodic markers, exploiting an alignment marker design with an increased number of features with a known arrangement, e.g. a periodicity, can help to improve the robustness and accuracy of the marker fitting.

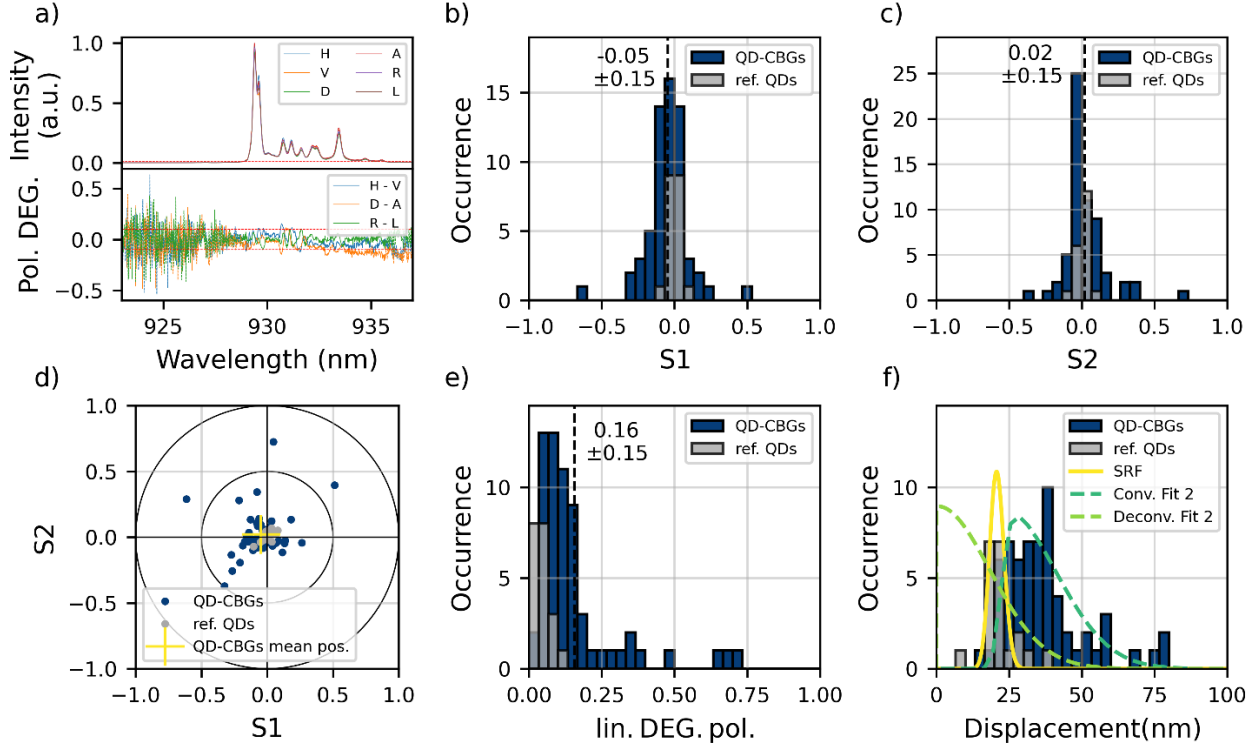

**Figure 3.: (a)** Spectra of a representative device for different polarizations (top) and the corresponding Stokes parameters for all wavelengths (bottom). **(b)**, and **(c)** Stokes parameters S1 and S2 of the CBGs and quantum dots in the membrane without cavity as reference measurements. The Stokes parameters are calculated by summing up the intensities above the threshold (red line in (a) top) first. **(d)** Linear polarization of the devises plotted in S1 vs S2. The devises show a small offset towards negative S1. **(e)** linear degree of polarization calculated from S1 and S2. The linear degree of polarization is translated to displacement **(f)** for each devise. The references QDs are treated as the system response function (SRF) and fitted with a Nakagami distribution, yielding a mean displacement of $(22.2 \pm 5.9)$ nm. The CBGs exhibit a mean displacement of $(37 \pm 14)$ nm without fitting. They are then fitted with a convolution of the SRF and a Nakagami distribution. The corrected, deconvoluted Nakagami distribution gives a mean displacement of $(15 \pm 11)$ nm demonstration the accuracy of the device placement.

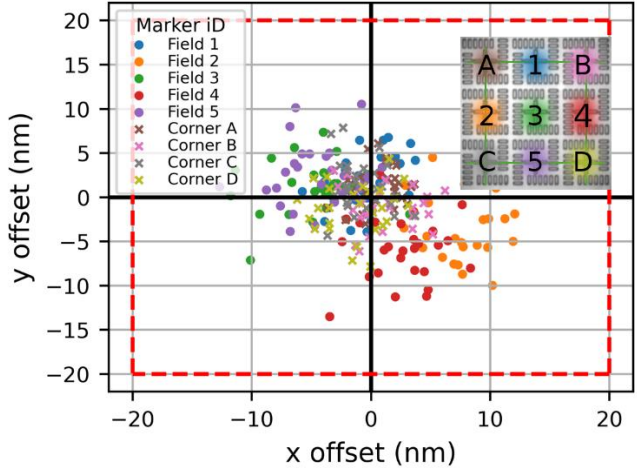

**Figure 4.:** Evaluation of the marker test fields (inset). The positions of the nine markers in each test field are determined. The four corner markers are used to define a local coordinate system, and the offsets between the measured marker positions and their intended positions (as specified in the layout) are calculated and presented. The average offset across all markers is (5.12 ±3.08) nm, indicating the accuracy of the established coordinate system.

### 3.2. Imaging Accuracy

The accuracies of marker and QD fitting alone are insufficient to reliably estimate the precision with which the cavities can be positioned around the QDs. Image distortions introduce additional errors [27]. To test the imaging accuracy inside a coordinate system, which is set up by the markers, we use the marker test field described in Section 3.1 (see Fig 4. inset). The marker test field is considered as two different sets of markers: the corner markers (Fig 4. A-D), which span our reference coordinate system, and the field markers (Fig 4. 1-5)**,** which are the remaining five. To assess the imaging accuracy, we acquire reflection images of 30 test fields and apply the image correction on each. For both the four corner markers as well as for the five additional markers, the offset between their nominal position given by the sample layout and the fitted position is calculated and plotted, see Fig 4. We observe two different sets of inaccuracies: First, the corner markers scatter around the origin, which leads to an imperfect coordinate system adding uncertainties to the QDs localized within it. Second, the markers inside the field have some systematic offsets from zero depending on their position inside the field. This suggests incomplete correction of image distortion, also adding to the total QDs localization accuracy. The coordinate system accuracy can be calculated as the mean of all markers, corner and field, and is (5.12 ± 3.08) nm, which includes the marker fitting accuracy from above. This value reflects the accuracy and precision of the established coordinate system used to determine the QD positions, taking into account errors arising from image distortions and marker fitting.

The QDs are fitted with a 2D-Gaussian function. The QD fit error is given by the standard deviation of the fit parameters and amounts to (4.4 ±3.7) nm for 155 fitted QDs. Two primary factors that can increase this error are low QD photoluminescence intensity and close proximity to neighboring

QDs, both of which reduce the reliability of the localization fit. These error sources, the imperfect coordinate system, together with not completely corrected distortions, both including the marker fitting accuracy and the QD fitting uncertainty, together with the transfer of the marker position between the reflection- and photoluminescence-image (See section 3.1), adds up to (9.1 ± 2.5) nm. This shows that photoluminescence-imaging is a viable option to localize QDs with respect to the markers that can be used for device fabrication. Note that for the total device positioning accuracy additional error sources have to be considered, e.g., the EBL overlay accuracy and the EBL writing accuracy. The total device placement accuracy is addressed in the next section.

### 3.3. Device positioning accuracy

Beside the QD localization with respect to the markers, fabrication of the CBGs afterwards and the marker deposition beforehand are additional sources for displacement. To test the total accuracy, we measured the polarizations of 87 devices and translated them to displacements (see section 2.5). Fig 3 a) presents the polarization-resolved spectra of a representative device, along with the corresponding Stokes parameters, which are calculated separately for each wavelength. Due to imperfections during the device process some of the CBGs showed significant mode splitting. Since mode splitting can also lead to polarized emission when the QD is spectrally aligned more with one mode than the other, we exclude devices with a mode splitting exceeding 2.5 nm (25) from our analysis. In Fig. 3 b)-f) the Stokes Parameters of the remaining 62 devices are shown. As comparison 20 QDs outside of CBGs are measured, for which we expect zero for all measured Stokes parameters. These 20 QDs are referred to as reference QDs. Most of the devices have a polarization of S1 and S2 smaller than 25% with small tendencies towards negative S1 (Fig 3 b)-f)). This hints a small systematic error, which could come from a systematic displacement in the order of 23 nm, or it could come from the fact that although the mode splitting of the cavities is small, it is almost never exactly zero and almost all the measured signal from the QDs is slightly blue detuned with respect to the cavity mode and is therefore spectrally overlapping more with the vertical mode of the cavity. However, since we cannot further quantify this effect, we assume that the measured polarization comes from device displacement and include the offset into our analysis. The mean of the Stokes parameter S1 (S2) is -0.05 ±0.15 (0.02 ±0.15) showing small average polarization but noticeable spread. The spread is an order of magnitude higher than for the reference QDs although the signal is at least an order of magnitude brighter. This indicates a random displacement of the devices with a spread around zero displacement. The Stokes parameter S3 shows a close to zero value of 0.026 ± 0.049 (-0.001 ± 0.023 for the reference QDs) and is therefore not shown. The absence of circular polarization is expected [29] and confirms the reliability of the measurement procedure. For each device the linear degree of polarization is calculated from S1 and S2, see Fig 3 e) and then translated to displacement using the simulations from Ref. [29]. The result is presented in Fig 3. f). As described in section 2.5, the displacement distribution is fitted with a convolution of the reference data and a Nakagami distribution. This results in an average displacement of (15 ± 11) nm. This can be interpreted as a total cavity placement accuracy of 15 nm with a precision of 11 nm. However, due to the limited reference dataset (20 QDs) and the dependence of the result on the exact SRF position, these values should be interpreted with caution. To give the complete picture we also report the raw values obtained directly from the data by fitting a Nakagami distribution without any correction, yielding a displacement of (33.5 ± 9.9) nm.

The raw (corrected) values are about a factor of 4 (2) higher compared to the evaluation of the marker test fields (section 3.2). This could come from the fact that the bare imaging evaluation does not include the displacement errors introduced by the CBG fabrication process with the marker-based EBL. Furthermore, for imaging the sample, a camera with an InGaAs sensor was used as the setup is used for QDs from 780 nm up to Telecom C-band 1550 nm QDs. This camera exhibits a significant number of hot pixels with their occurrence increasing as the integration time lengthens. To balance signal quality and the occurrence of hot pixels, an integration time of 1500 ms is chosen, during which approximately 2.1% of the pixels exhibit abnormally high intensity values. A longer integration time would improve the signal-to-noise ratio and enhance QD localization precision but would also increase the number of hot pixels. Although correction algorithms are used to mitigate their effect, the information of these pixels is inevitably lost. Also, as mentioned before, a splitting of the cavity mode can introduce polarized emission [34]. This can lead to an overestimation of the displacement determined from the polarization.

Compared to previously reported QD localization and positioning techniques, our method shows improved accuracy relative to previous photoluminescence imaging approaches. Although the achieved accuracy is comparable to that of cathodoluminescence imaging and in-situ electron beam lithography, the precision attained in our method surpasses all previously reported techniques [24], resulting in an overall high device yield of 68%, if a residual polarization of 10% and a reduction of the Purcell effect by 20% is acceptable. To calculate the device yield for polarization values below 10%, the linear degree of polarization was fitted with a convolution of a Nakagami distribution and the SRF, similar to the approach used in the displacement evaluation.

## 4. Conclusions

In summary, we introduced a periodic alignment marker design and achieved a marker localization accuracy of $(2.87 \pm 0.65)$ nm. We believe that these or similar periodic markers can contribute to improved QD imaging within the research community. By arranging these markers in a grid with a spacing of half the field size, we investigated the total QD localization accuracy within a coordinate system defined by the markers. Considering QD and marker localization uncertainties, coordinate system imperfections, and marker position transfer errors between reflection and photoluminescence images, we determined a total QD localization uncertainty of $(9.1 \pm 2.5)$ nm.

To assess the final integration accuracy, we processed 87 CBGs and evaluated their positioning by measuring the device polarization. With a mean displacement of $(33.5 \pm 9.9)$ nm (raw) and $(15 \pm 11)$ nm (corrected), our approach reaches the accuracy of cathodoluminescence imaging or in-situ e-beam lithography, and it significantly improves the precision. While in-situ e-beam lithography integrates low-temperature cathodoluminescence and e-beam lithography within a single setup, thereby increasing system complexity, marker-based photoluminescence imaging relies on two separate setups that can be individually optimized for their respective purposes. We believe that the photoluminescence imaging accuracy can be improved further by reducing the setup vibrations; replacing the InGaAs sensor with a scientific-grade CMOS camera, which would offer a better signal-to-noise ratio and fewer hot pixels, however, only in the near infrared and not in the telecom wavelength range; or improving image correction methods like e.g. using Zernike polynomials [27]. Additionally, uncertainties from the marker-based e-beam lithography opens room for further investigation and improvements.

Our findings demonstrate that photoluminescence imaging is an effective method to enable the deterministic fabrication of nanophotonic devices specifically tailored to the optical emitter properties, and whose performance is sensitive to small displacements, as seen in circular Bragg gratings. With a device yield of 68%, this method ensures the reliable fabrication of high-quality devices across each sample.

## Declaration

**Acknowledgements**. We would like to express our sincere gratitude to Silke Kuhn for her invaluable assistance in sample fabrication and to Johannes Michl for his excellent work in creating the graphical abstract.

**Availability of data and material.** The datasets used and/or analyzed during the current study are available from the corresponding author on reasonable request.

**Competing interests.** The authors declare that they have no competing interests.

**Funding.** The authors acknowledge the support of the state of Bavaria and the German Ministry for Research and Education (BMBF) within the projects QR.X (FKZ: 16KISQ010), QR.N (FKZ: 16KIS2209), PhotonQ (FKZ: 13N15759), Qecs (FKZ: 13N16272) and QD-E-QKD (FKZ: 16KIS1672K).   We also acknowledge financial support by the German research foundation (DFG) under the project reference DIP FI947/6-1.

**Authors' contributions.** QB and CK performed the experiments, measured the data and analyzed the data. QB and CK wrote the software for measuring and evaluating the imaging data. QB compiled the data and wrote the manuscript draft. AZ helped writing the code of the marker fitting routine. GP helped to develop the idea of periodic features for markers. MH wrote the software for the polarization resolved measurement. YR did simulation for the dependence between displacement and polarization in CBGs. AP SH THL supervised the work and reviewed and edited the manuscript. All authors commented and revised the manuscript and approved the final manuscript.